\documentclass[10pt, conference]{IEEEtran}
\IEEEoverridecommandlockouts
\usepackage{cite}

\usepackage{amsmath,amssymb,amsfonts,mathtools, bm}
\usepackage{amsthm}
\usepackage{algorithm}
\usepackage{algorithmic}
\usepackage{graphicx}
\usepackage{textcomp}
\usepackage{xcolor,float}
\usepackage{tabularx}
\usepackage{textcomp}
\usepackage{diagbox}
\usepackage{hyperref}
\usepackage{booktabs}
\usepackage{subfigure}
\newcounter{MYtempeqncnt}

\newtheorem{problem}{Problem}

\allowdisplaybreaks
\renewcommand{\baselinestretch}{1}

\begin{document}

\title{Scalable Resource Management for Dynamic MEC: An Unsupervised Link-Output Graph Neural Network Approach}
\author{
\IEEEauthorblockN{
Xiucheng Wang\IEEEauthorrefmark{1},
Nan Cheng\IEEEauthorrefmark{1},
Lianhao Fu\IEEEauthorrefmark{1},
Wei Quan\IEEEauthorrefmark{2},
Ruijin Sun\IEEEauthorrefmark{1},
Yilong Hui\IEEEauthorrefmark{1},\\
Tom Luan\IEEEauthorrefmark{3},
Xuemin (Sherman) Shen\IEEEauthorrefmark{4}\\
}
\IEEEauthorblockA{
\IEEEauthorrefmark{1}School of Telecommunications Engineering, Xidian University, Xi'an, 710071, China\\
\IEEEauthorrefmark{2}School of Electronic and Information Engineering, Beijing Jiaotong University, Beijing 100044, China\\
\IEEEauthorrefmark{3}School of Cyber Engineering,
Xidian University, Xi'an, 710071, China\\
\IEEEauthorrefmark{4} Department of Electrical and Computer Engineering, University of Waterloo, Waterloo, ON N2L 3G1, Canada\\
Email: \{xcwang\_1, lhfu\}@stu.xidian.edu.cn, dr.nan.cheng, weiquan@bjtu.edu.cn,\\ \{sunruijin, ylhui, tom.luan\}@xidian.edu.cn, sshen@uwaterloo.ca}}
    \maketitle

\IEEEdisplaynontitleabstractindextext

\IEEEpeerreviewmaketitle

\begin{abstract}
Deep learning has been successfully adopted in mobile edge computing (MEC) to optimize task offloading and resource allocation. However, the dynamics of edge networks raise two challenges in neural network (NN)-based optimization methods: low scalability and high training costs. Although conventional node-output graph neural networks (GNN) can extract features of edge nodes when the network scales, they fail to handle a new scalability issue whereas the dimension of the decision space may change as the network scales. To address the issue, in this paper, a novel link-output GNN (LOGNN)-based resource management approach is proposed to flexibly optimize the resource allocation in MEC for an arbitrary number of edge nodes with extremely low algorithm inference delay. Moreover, a label-free unsupervised method is applied to train the LOGNN efficiently, where the gradient of edge tasks processing delay with respect to the LOGNN parameters is derived explicitly. In addition, a theoretical analysis of the scalability of the node-output GNN and link-output GNN is performed. Simulation results show that the proposed LOGNN can efficiently optimize the MEC resource allocation problem in a scalable way, with an arbitrary number of servers and users. In addition, the proposed unsupervised training method has better convergence performance and speed than supervised learning and reinforcement learning-based training methods. The code is available at \url{https://github.com/UNIC-Lab/LOGNN}.
\end{abstract}


\section{Introduction}
In recent years, mobile edge computing (MEC) has garnered widespread attention for its ability to reduce task processing latency by providing edge computing resources to users with limited computing capabilities \cite{7488250}. The conventional approach involves transmitting user tasks to the server via a wireless network, with several variables optimized to minimize task computing delays, such as offloading proportion, user transmission power, and server computing resource allocation, among others. Given the critical role of MEC in enhancing user experience, numerous researchers have focused their efforts on designing resource management methods that can further optimize MEC performance \cite{8672604,9369456}. Drawing inspiration from the remarkable achievements of deep learning, several studies have employed neural networks (NN) to address the optimization problem in MEC, outperforming traditional optimization methods both in terms of performance and algorithm execution time \cite{9860495,8976180}.

The dynamics of MEC systems, which are changes in numbers and locations of the edge servers (e.g., flying drones as MEC servers) and users (e.g., vehicular users), raise two challenges in NN-based methods: low scalability and high training costs \cite{shen2023toward}. Scalability plays a crucial role in determining whether a new NN architecture needs to be manually designed and retrained when the number of edge servers and users undergoes changes. Additionally, training costs are a key factor in determining the ability of an NN-based method to be rapidly deployed in the edge network. Some recent works explore using graph neural network (GNN) to deal with the scalability issue in input space, which includes usually the features of network nodes. This is because the inference of GNN relies on the message passing method which is an input dimension-independent method \cite{shen2020graph,wang2022joint}. However, in MEC, the dimension of offloading and resource allocation decisions can also change as the network scales. For instance, a task can be offloaded to a varying number of servers, thereby leading to a unique scalability issue, i.e., scalability in the decision space. Unfortunately, the conventional node-output GNN, which allocates resources from an edge node to others through the dimension-fixed node features vector, fails to address the dimension change of the decision space caused by the changing size of the edge network. Thus, the development of a novel approach that can efficiently tackle the scalability issues in both input and decision space remains a crucial area of research in optimizing MEC.

This paper presents a novel resource management scheme for dynamic MEC that leverages the capabilities of a link-output Graph Neural Network (LOGNN). Despite the fixed dimension of output link features, the number of graph links increases as the number of edge servers and users increases. Therefore, by defining the proportion of resources allocated from the servers/users to each user/server as the link feature from server/user nodes to user/server node, the proposed LOGNN-based resource management scheme can efficiently deal with the scalability issue in decision space. In addition to scalability, high training costs are another essential drawback of NN-based methods in dynamic MEC. The high training cost comprises two main parts: the extreme costs of obtaining optimal solutions as training labels and the low convergence speed. Although the reinforcement learning (RL)-based training method alleviates the dependency on training labels \cite{8650160}, its convergence speed and performance cannot be guaranteed. To overcome these challenges, this paper exploits a label-free unsupervised method to train the LOGNN. Parameters in the LOGNN are updated by the gradient of edge task processing delays with respect to the LOGNN parameters, where the task processing delays are derived explicitly from the output of the LOGNN. This approach not only reduces the training cost but also improves the convergence speed of the LOGNN. The main contributions of the paper are summarized as follows.

\begin{enumerate}
    \item An innovative resource management scheme based on LOGNN is presented that efficiently addresses the scalability issues in both input and decision space, as the number of edge servers and users in dynamic MEC systems changes. The LOGNN is theoretically analyzed and compared to conventional node-output GNNs, demonstrating its superior adaptability.
    \item A label-free unsupervised training method for LOGNN is introduced, which leverages the gradient of task processing delays concerning the GNN parameters as a loss function, accelerating the convergence speed and enhancing overall performance.
    \item Simulation results show combining the proposed LOGNN and the unsupervised training method, can flexibly and efficiently optimize task offloading and resource allocation in MEC. Furthermore, it outperforms supervised learning and RL-based training methods in terms of convergence speed and performance, showcasing its potential for practical implementation in dynamic MEC networks.
\end{enumerate}

\section{Graph Modeling of MEC Resources Management}
\begin{figure}[ht]
  \centering
  \includegraphics[width=0.75\columnwidth]{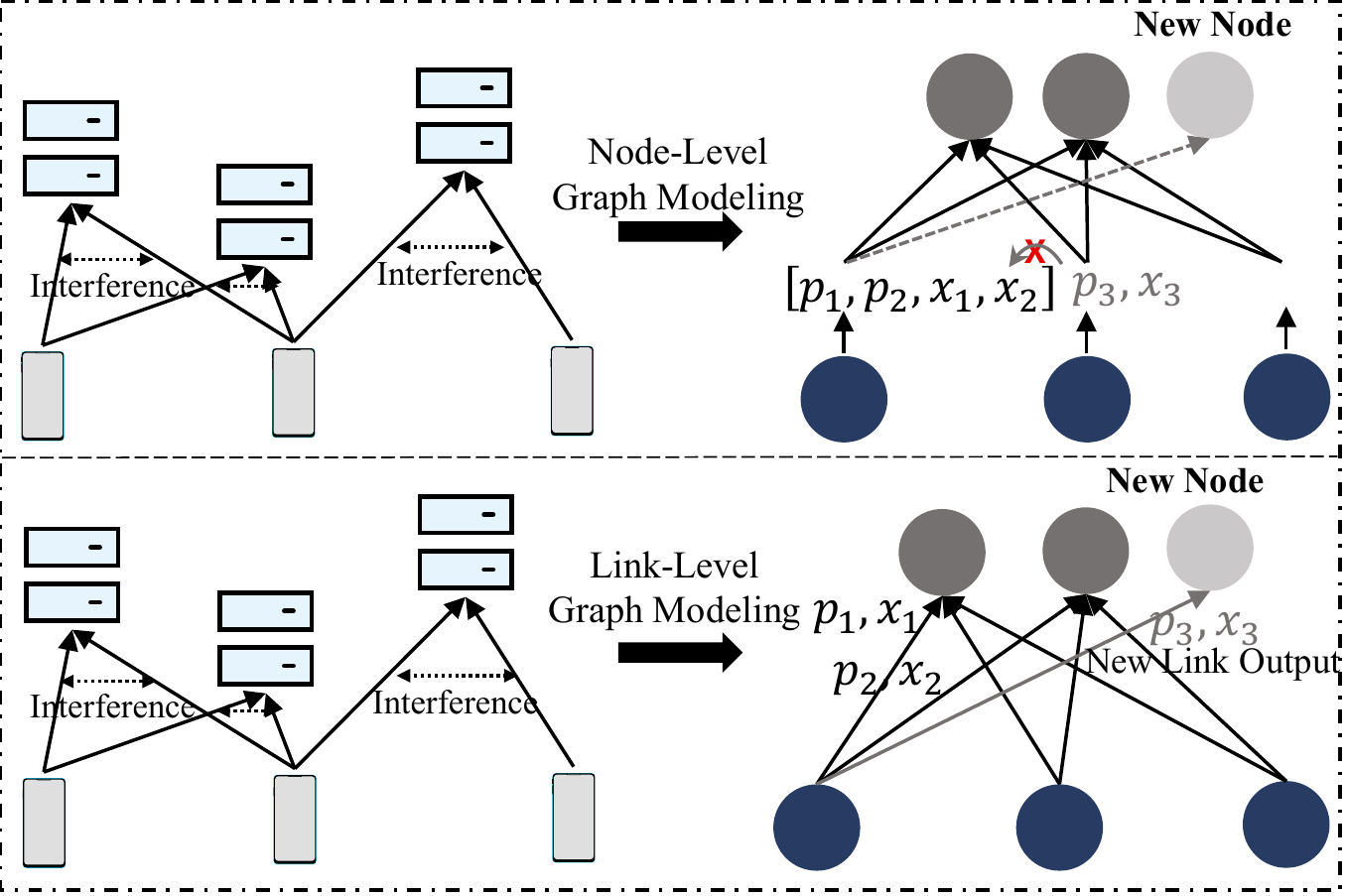}
   \vspace{-9pt}
  \centering \caption{Illustration of graph modeling of the MEC resources management for node-output and link-output GNNs.} 
  \label{system}
   \vspace{-9pt}
\end{figure}
\subsection{System Model and Problem Formulation}
In this paper, we consider the scenario with $N$ users randomly located in the plane, and for each user $i,i\in\{1,2,\cdots, N\}$, there is a task with size $d_i$ to be processed. Meanwhile, $M$ servers are also randomly located in the plane with computing resources $f_{j}^{s}$ for server $j$; since the computing resources of users are limited, all tasks are transmitted to servers to process. To balance the workloads among servers and reduce task computing latency, similar to \cite{apostolopoulos2020risk}, each task can be divided into multiple parts\footnote{We call the parts of a task as subtasks in this paper.} with different proportions and transmitted to different servers. Each server uses an orthogonal frequency to communicate with users. Thus, the transmission rate between user $i$ and server $j$ is
\begin{align}
    &r_{i,j} = b\log_2 \left(1+\frac{p_{i,j}h_{i,j}}{\sum_{k\neq i}^{N}p_{k,j}h_{k,j}+\sigma^2}\right),\label{rate}
\end{align}
where $b$ is the bandwidth, $p_{i,j}$ is the transmit power for user $i$ to transmit a subtask to server $j$,  $\sigma^2$ is the noise power, $h_{i,j}$ is the channel gain between user $i$ and server $j$, and $g_0$, $l_0$ are the reference channel gain and distance, respectively.

Since different users transmit subtasks with various sizes and transmission rates, to fully exploit the computing resources of edge servers and reduce the computing delay, servers need to determine the amount of computing resources to process a special subtask, and the server computing latency can be calculated as
\begin{align}
    T_{i,j}^{com}= \frac{x_{i,j}d_ic}{f_{i,j}},
\end{align}
where $x_{i,j}$ is the offloading proportion of a subtask from user $i$ to server $j$, $c$ is the constant computing factor that determines the amount of central processing unit cycles to compute one-bit data, and $f_{i,j}$ is the computing resource allocated to process the subtask offloaded from user $i$ to server $j$.

Therefore, the task computing delay minimization problem is formulated as
\begin{problem}\label{p-1}
\begin{align}
    &\min_{\mathbf{x,p,f}}\sum_{i=1}^{N}\sum_{j=1}^{M} \frac{d_{i}x_{i,j}}{b\log_2 \left(1+\frac{p_{i,j}h_{i,j}}{\sum_{k\neq i}^{N}p_{k,j}h_{k,j}+\sigma^2}\right)}+\frac{x_{i,j}d_ic}{f_{i,j}},\label{obj}
  \end{align}
  \begin{align*}
    &s.t. \quad x_{i,j}\geq 0,\;\;\forall i\in\{1,\cdots,N\}\wedge\forall j\in\{1,\cdots,M\},\tag{3a}\label{c1}\\
    &\qquad\;\, p_{i,j}\geq 0,\;\;\forall i\in\{1,\cdots,N\}\wedge\forall j\in\{1,\cdots,M\},\tag{3b}\label{c2}\\
    &\qquad\;\, f_{i,j}\geq 0,\;\;\forall i\in\{1,\cdots,N\}\wedge\forall j\in\{1,\cdots,M\},\tag{3c}\label{c3}\\
    &\qquad\;\, \sum_{j=1}^{M} x_{i,j}= 1,\qquad\qquad\qquad\;\,\,\,\,\forall i\in\{1,\cdots,N\},\tag{3d}\label{c4}\\
    &\qquad\;\, \sum_{j=1}^{M} p_{i,j}\leq p_{max},\qquad\qquad\quad\;\,\forall i\in\{1,\cdots,N\},\tag{3e}\label{c5}\\
    &\qquad\;\, \sum_{i=1}^{N} f_{i,j}\leq f_{j}^{s},\qquad\qquad\qquad\;\,\forall j\in\{1,\cdots,M\},\tag{3f}\label{c6}
\end{align*}
\end{problem}
\noindent where the objective function \eqref{obj} is to minimize the sum of transmission delay and server computing latency for all user tasks. \eqref{c1}-\eqref{c3} constrain the task offloading proportion, transmitting power, and the computing resources allocated to users must be larger than or equal to zero. \eqref{c4} guarantees that a task is fully transmitted to the servers to process, and \eqref{c5} constraints that the user can allocate the transmission power for all servers, but the sum of transmission power cannot exceed the maximum power $p_{max}$.  \eqref{c6} guarantees that for each server the sum of computing resources allocated to all users cannot be larger than its maximum resources.

\subsection{Graph Modeling of MEC System}
To effectively reduce total tasks processing delay, Problem \ref{p-1} is modeled as a graph link weights regression problem, where users and servers are modeled as graph nodes and wireless channels between users and servers are modeled as graph links. The node feature matrix $\mathbf{Z}=\{\mathbf{Z}^{u},\mathbf{Z}^{s}\}$ consists of user node feature matrix $\mathbf{Z}^{u}\in \mathbb{R}^{N\times1}$ which is given by $\mathbf{Z}^{u}_{(i,:)}=[d_i]$, and the server node feature matrix $\mathbf{Z}^{s}\in \mathbb{R}^{M\times1}$ which is given by $\mathbf{Z}^{s}_{(j,:)}=[f_j^s]$. Since in Problem \ref{p-1}, users need to allocate subtask sizes and transmit power to servers, and meanwhile, servers need to allocate computing resources to users, for each pair of user and server, the resource allocation is bi-directional. As a consequence, the problem graph is modeled as a bi-direction graph where the adjacency feature array $\mathbf{A}\in \mathbb{R}^{(N+M)\times (N+M)}$ is given as
\begin{align}
    \mathbf{A}_{(i,j,:)} =\begin{cases}
    h_{i,j} &\left(i\in\mathcal{V}_{u}\wedge j\in\mathcal{V}_{s}\right)\vee\left(i\in\mathcal{V}_{s}\wedge j\in\mathcal{V}_{u}\right),\\
    0 &otherwise,
    \end{cases}\notag
\end{align}
where $\mathcal{V}_u$ and $\mathcal{V}_s$ are node sets of users and servers.

As illustrated in Fig.~\ref{system}, the optimization variables in Problem \ref{p-1} are modeled as the link weights $\bm{\mathcal{E}}=\{\bm{\mathcal{E}}^{u},\bm{\mathcal{E}}^{s}\}$, where $\bm{\mathcal{E}}^{u}\in \mathbb{R}^{N\times M\times2}$ is given by $\bm{\mathcal{E}}^{u}_{(i,:,:)}=\left[[x_{i,1},p_{i,1}],[x_{i,2},p_{i,2}],\cdots,[x_{i,M},p_{i,M}]\right]$ and $\bm{\mathcal{E}}^{s}\in \mathbb{R}^{M\times N}$ is given by $\bm{\mathcal{E}}^{s}_{(j,:)}=[f_{1,j},f_{2,j},\cdots,f_{N,j}]$. With the above notations, Problem \ref{p-1} can be rewritten as
\begin{problem}\label{p-2}
\begin{align}
    &\min_{\bm{\mathcal{E}}} \sum_{i=1}^{N}\sum_{j=1}^{M}\frac{\mathbf{Z}^{u}_{(j,2)}\bm{\mathcal{E}}^{u}_{(i,j,0)}}{b\log\left(1+\frac{\bm{\mathcal{E}}^{u}_{(i,j,1)}\mathbf{A}_{(i,j)}}{\sum_{k\neq i}^{N}\bm{\mathcal{E}}^{u}_{(k,j,1)}\mathbf{A}_{(k,j)}+\sigma^2}\right)}\notag\\
    &\qquad\qquad\quad+\frac{\mathbf{Z}^{u}_{(j,2)}\bm{\mathcal{E}}^{u}_{(i,j,0)}c}{\bm{\mathcal{E}}^{s}_{(j,i)}},\label{obj-2}
\end{align}
\begin{align*}
    &s.t. \quad \bm{\mathcal{E}}\succeq\mathbf{0},\tag{4a}\label{c2-1}\\
    &\qquad \sum_{j=1}^{M}\bm{\mathcal{E}}^{u}_{(i,j,0)}=1,\quad\qquad\qquad\;\,\forall i\in\{1,\cdots,N\},\tag{4b}\label{c2-2}\\
    &\qquad \sum_{j=1}^{M}\bm{\mathcal{E}}^{u}_{(i,j,1)}\leq p_{max},\quad\qquad\;\;\,\forall i\in\{1,\cdots,N\},\tag{4c}\label{c2-3}\\
    &\qquad \sum_{i=1}^{N}\bm{\mathcal{E}}^{s}_{(j,i)}\leq f_{j}^{s},\quad\qquad\qquad\;\forall j\in\{1,\cdots,M\},\tag{4d}\label{c2-4}
\end{align*}
\end{problem}
\noindent where \eqref{c2-1} constrains $x$, $p$, $f$ larger than 0, which is equal to constraints \eqref{c1}-\eqref{c3}, and \eqref{c2-2}, \eqref{c2-3}, \eqref{c2-4} correspond to the \eqref{c4}, \eqref{c5}, \eqref{c6} in Problem \ref{p-1}, which have same constraint condition.

\section{GNN Based Link Weight Regression for MEC Resource Allocation}
\subsection{Structure of Proposed GNN}
In order to effectively extract features of MEC, the proposed LOGNN $\mathcal{G}$ is based on the architecture of graph attention network (GAT), where for a specific node the influence of its different neighbor node is calculated as attention value \cite{velickovic2018graph}. The details of the proposed LOGNN method are 
\begin{align}
    &m_{i,j} =\text{AGG}(\mathbf{Z}_{i},\mathbf{Z}_{j}),\,\quad\quad\quad  \mathbf{A}_{i,j}\neq0, \label{u-message}\\
    &\varsigma_i\;=\phi(\mathbf{Z}_i,\rho\{\alpha_{i,j}m_{i,j}:j\in\mathcal{N}(i)\},\label{ugnn-agg}\\
    &\alpha_{i,j}\;\;=\frac{\exp\left(\text{LeakyReLU}\left(\left[W_1\varsigma_{i}||W_2\varsigma_{j}\right]\right)\right)}{\sum_{k\in \mathcal{N}(i)}\exp\left(\text{LeakyReLU}\left(\left[W_1\varsigma_{i}||W_2\varsigma_{k}\right]\right)\right)},\label{u-att}\\
    &[p_{i,j},x_{i,j}]=W_3[\varsigma_{i},\varsigma_{j}],\quad\qquad\quad\;\;\;  i \in \mathcal{V}^{u}\wedge j \in \mathcal{V}^{s},\label{u-out}\\
    &f_{i,j}=W_4[\varsigma_{j},\varsigma_{i}],\quad\qquad\qquad\quad\;\;\;\;\;  i \in \mathcal{V}^{u}\wedge j \in \mathcal{V}^{s},\label{s-out}
\end{align}
where $\mathcal{V}^{u}$ and $\mathcal{V}^{s}$ are the node set of users and servers, and $\mathcal{N}(i)$ is the set of adjacency nodes for node $i$, and LeakyReLU$(x)$ is equal to $x$ when $x$ is larger than 0, otherwise is $\epsilon x$ that $\epsilon$ is a factor larger than 0 but smaller than 1. $m_{i,j}$ is the extracted message passed from neighbor node $i$ to node $j$, and AGG$(\cdot)$ is trainable aggregation net to extract features of neighbor nodes. $\rho(\cdot)$ is a message concatenate function used to compress the message from all adjacency nodes to a vector, which is usually set as $\max (\cdot)$ or mean$(\cdot)$, $a_{i,j}$ is the attention value to evaluate the influence of adjacency node $j$ on the current node $i$, and $\phi$ is a trainable neural network which is used to update node features from the initial input features $Z_i$ to output features $\varsigma_i$. $W_1$, $W_2$, $W_3$ and $W_4$ are matrices of trainable parameters. $W_1$ and $W_2$ are used to calculate the attention value $a_{i,j}$, $W_3$ is used to determine the link weight $p_{i,j}$ and $x_{i,j}$ from user $i$ to server $j$, and $W_4$ is used to determine the link weight $f_{i,j}$ from server $j$ to user $j$.

\subsection{Unsupervised Learning-Based Efficient Training Method}
Since the objective function \eqref{obj-2} is differentiable, we exploit an unsupervised learning-based training method for LOGNN, where the gradient of the objective function of the equation \eqref{obj} $G(\bm{\mathcal{E}}|\mathbf{Z}^{u},\mathbf{Z}^{s},\mathbf{A})$ with respect to parameters $\bm{\theta}$ of LOGNN can be calculated by chain rule as equation \eqref{gra}, and thus the parameters $\bm{\theta}$ is updated as \eqref{update}, where $lr$ is the learning rate.
\begin{figure*}[ht]
\normalsize
\setcounter{MYtempeqncnt}{\value{equation}}
\begin{align}
    \nabla_{\bm{\theta}} G(\bm{\mathcal{E}}|\mathbf{Z}^{u},\mathbf{Z}^{s},\mathbf{A})&=\nabla_{\bm{\mathcal{E}}} G(\bm{\mathcal{E}}|\mathbf{Z}^{u},\mathbf{Z}^{s},\mathbf{A})|_{\bm{\mathcal{E}}=\mathcal{G}_{\bm{\theta}}(\mathbf{Z}^{u},\mathbf{Z}^{s},\mathbf{A}|\bm{\theta})}\nabla_{\bm{\theta}}\mathcal{G}_{\bm{\theta}}(\mathbf{Z}^{u},\mathbf{Z}^{s},\mathbf{A}|\bm{\theta}),\notag\\
    &=\nabla_{\mathbf{x,p,f}}\left(\sum_{i=1}^{N}\sum_{j=1}^{M} \frac{d_{i}x_{i,j}}{b\log_2 \left(1+\frac{p_{i,j}h_{i,j}}{\sum_{k\neq i}^{N}p_{k,j}h_{k,j}+\sigma^2}\right)}+\frac{x_{i,j}d_ic}{f_{i,j}}\right)\nabla_{\bm{\theta}}\mathcal{G}_{\bm{\theta}}(\mathbf{Z}^{u},\mathbf{Z}^{s},\mathbf{A}|\bm{\theta}),\label{gra}
\end{align}
\setcounter{equation}{\value{MYtempeqncnt}}
\hrulefill
\vspace*{4pt}
\end{figure*}
\begin{align*}
    \bm{\theta}=\bm{\theta}-lr \nabla_{\bm{\theta}}G(\bm{\mathcal{E}}|\mathbf{Z}^{u},\mathbf{Z}^{s},\mathbf{A})\tag{11}\label{update},
\end{align*}
Through such an unsupervised training method, the LOGNN is trained without optimal solutions as labels, which are challenging and costive to obtain for such a non-convex problem. Moreover, the proposed unsupervised training method outperforms the usually used actor-critic reinforcement training method in both convergence speed and performance \cite{8672604,8650160}.

\subsection{Theoretical Analysis of Proposed LOGNN Method}

Algorithmic scalability is paramount in determining whether a trained GNN can directly optimize resource allocation for dynamic MEC systems with fluctuating numbers of edge servers and users. However, employing differently sized GNNs and retraining them is impractical due to exorbitant retraining costs. The proposed LOGNN-based resource management method demonstrates superior scalability compared to conventional node-output GNN, as LOGNN retains adaptability in the decision space dimension, making it well-suited for dynamic MEC systems with varying edge servers and user counts.

In essence, the number of users for each server dictates the feasible decision space dimension for allocating computing resources. Given a server allocating $f$ computing resources to $n$ users, the feasible decision space resembles an $n$-dimensional cube with a side length of $f$, resulting in a space size of $f^{n}$. Conversely, the GNN aggregation net (comprising function $\phi(\cdot)$ in \eqref{ugnn-agg}) constrains each node's output features dimension to a fixed number $m$, which corresponds to the output layer's neural nodes count \cite{wu2020comprehensive}. Employing the Softmax function usually limits the sum of resources allocated to users, ensuring it does not exceed $f$. Thus, the GNN output space forms an $m$-dimensional cube with sides of length $f$, amounting to a space size of $f^{m}$. In situations where the user count exceeds $m$, the node-output GNN fails to identify the optimal solution, as the decision space dimension falls short of the feasible decision space. Likewise, the node-output GNN struggles to manage users' offloading and transmit power allocation when server numbers surpass the output layer's neural nodes count in the aggregation net.

As for LOGNN, output link features also present a fixed size. However, the link count increases alongside server and user numbers. By connecting servers to all users, the link-output GNN output space dimension matches the user count, allowing the link-output GNN to address resource allocation challenges for an arbitrary user count. Similarly, LOGNN facilitates users in making offloading and power allocation decisions with an arbitrary server count.

\section{Simulation Results and Discussion}
\begin{figure}[ht]
    \centering
    \subfigure[Convergence performance on training set.]{
        \includegraphics[width=0.75\columnwidth]{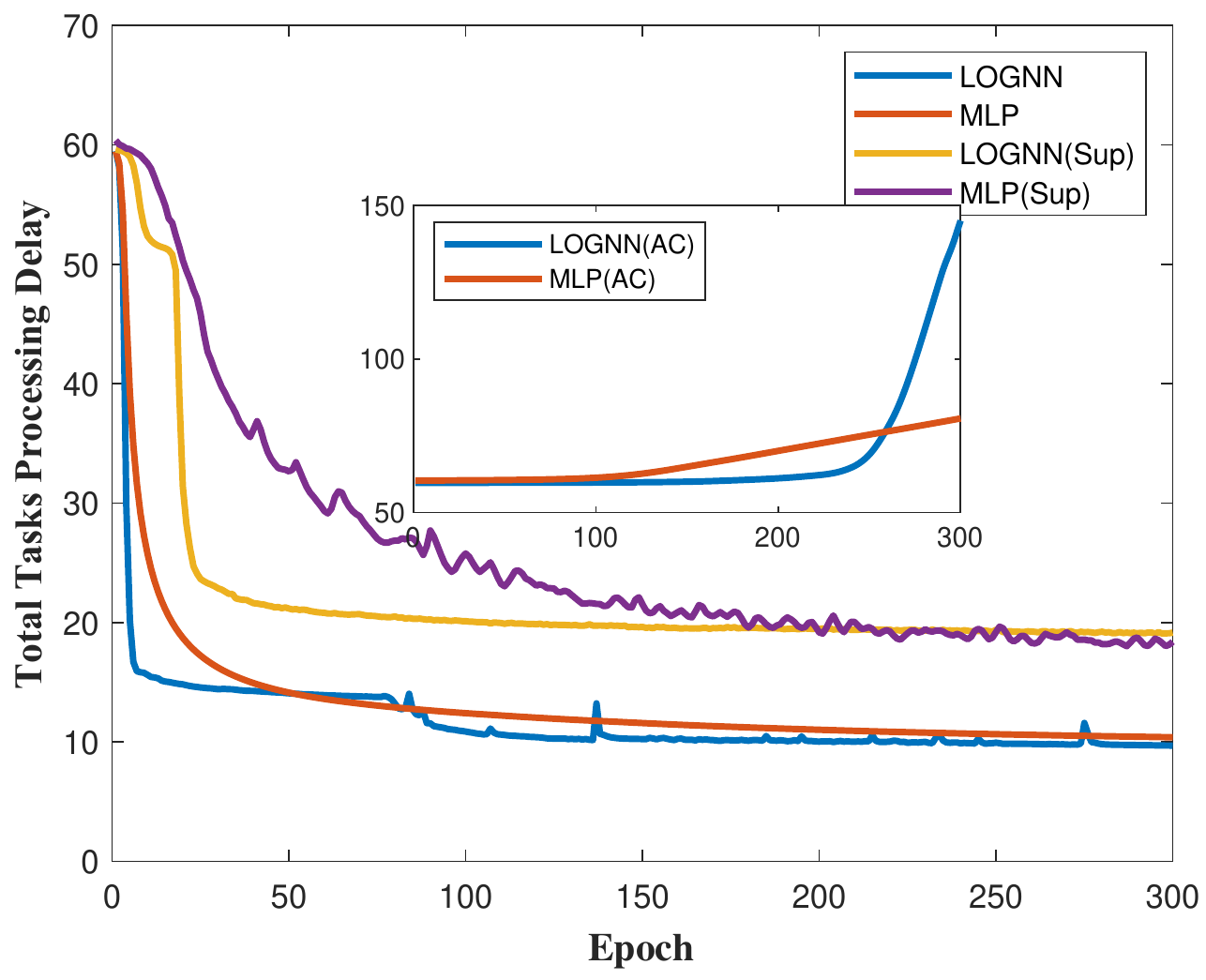}\label{train}
    }
    \subfigure[Convergence performance on test set.]{
        \includegraphics[width=0.75\columnwidth]{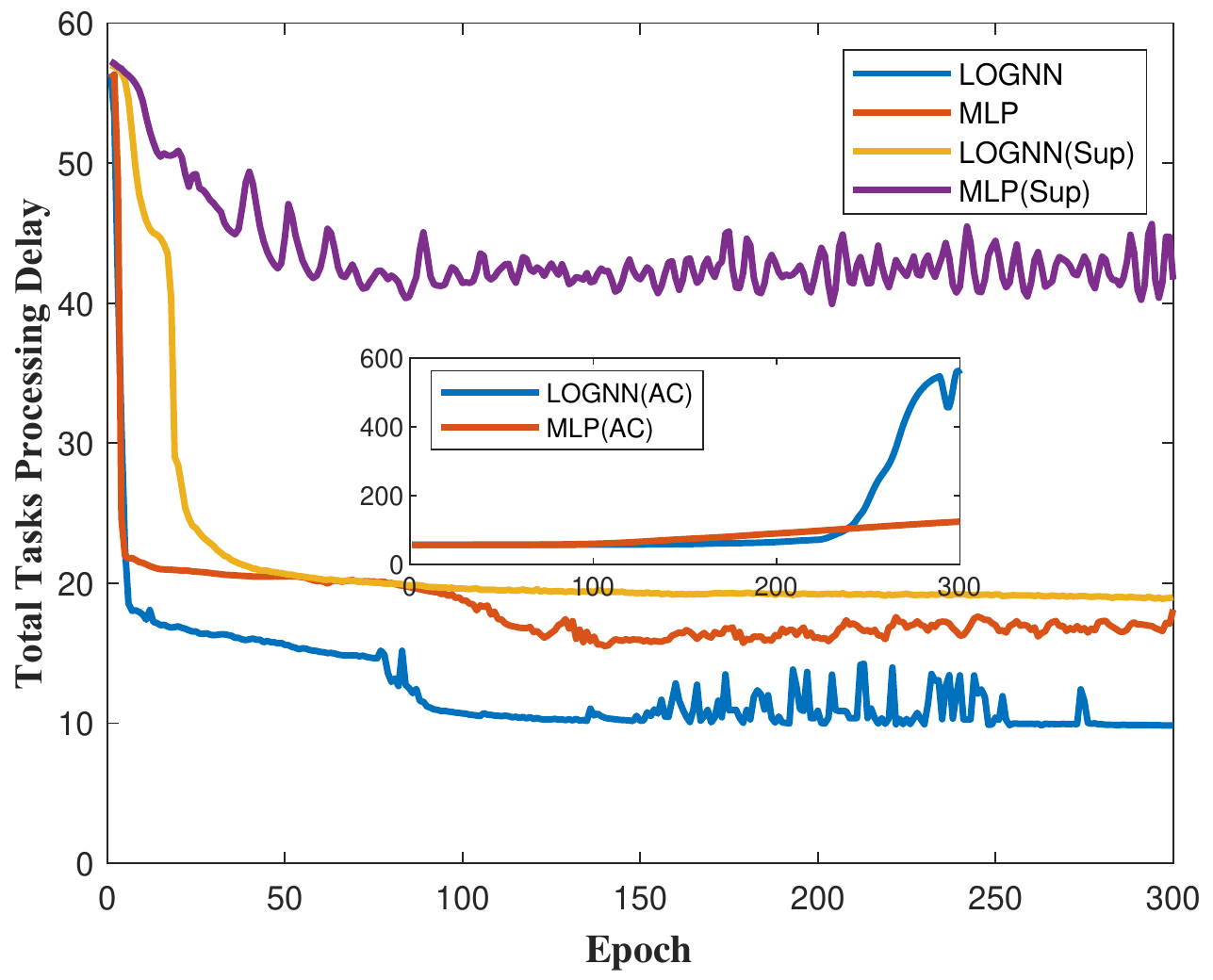}\label{test}
    }
    \caption{Convergence performance for different algorithms on different training methods.}
    \vspace{-20pt}
    \label{fig-con}
\end{figure}
\subsection{Simulation Settings}
In this part, simulation experiments are conducted to evaluate the performance of the proposed LOGNN-based MEC resource management method. Similarly to \cite{shen2020graph}, the channel gains are generated randomly with $h_{i,j}\sim{U(0,1)}$, the package size of users' tasks are randomly generated with $d_i\sim{U(0,1)}$, and the computing resources of servers are randomly generated with $f_{j}^{s}\sim{U(0,1)}$, where $U(\cdot,\cdot)$ is the uniform distribution, and $p_{max}$ is set to 1. We consider the following benchmarks for comparison.\\
$\bullet$MLP(DI): A multi-layer perceptron (MLP) proposed in \cite{liang2019towards} is used to extract features of MEC, which is pre-trained on a specific number of edge servers and users since the dimension of input and output for MLP is a fixed number. When the number of nodes in MEC exceeds the number of pre-trained nodes, the MLP can only optimize the performance of a subset of the nodes and randomly allocates $\mathbf{x,p,f}$ to other nodes.\\
$\bullet$MLP(TR): Whenever the number of nodes changes, a new MLP is trained and deployed to cope with the change in the dimension of the input-output space.\\
$\bullet$ GA: Genetic algorithm is an algorithm that uses the mechanism of biological evolution and is widely used to solve non-convex optimization problems \cite{li2020genetic}.\\
In the simulation, the proposed LOGNN is pre-trained using numerous data of different numbers and locations of edge servers and users. In the inference procedure, the pre-trained LOGNN is directly used to optimize the MEC task offloading and resource allocation  without any re-training. All algorithms are run on the graphic processing unit of the Tesla A100. More hyper-parameters of LOGNN and other compared algorithms are shown in Table \ref{params}.

\begin{table}[ht]
    \centering
    \setlength{\tabcolsep}{3mm}{}
    \caption{hyper-parameter setting of different algorithms}
    \resizebox{0.70\columnwidth}{!}{
    \begin{tabular}{c|ccc}
    \toprule[0.5pt]\toprule[0.5pt]
       Training Parameters  & LOGNN & MLP & GA\\ \hline 
       number of layers   & 2 & 4 & $\backslash$\\ \hline 
       hidden dimension & 64 & 64 & $\backslash$ \\ \hline 
       $lr$ & 1e-4 & 1e-4 & $\backslash$ \\ \hline 
       optimizer  & Adam & Adam & $\backslash$ \\ \hline 
       training epochs & 500 & 500 & 500 \\ \hline 
       number of training samples & 2048 & 2048 & $\backslash$ \\ \hline 
       batch size & 32 & 32 & $\backslash$ \\\hline 
       number of population &$\backslash$ & $\backslash$ & 200 \\\hline  
       retain rate &  $\backslash$ & $\backslash$ & 0.4 \\ \hline 
       mutation rate & $\backslash$ & $\backslash$ & 0.2 \\ \hline 
       selection rate & $\backslash$ & $\backslash$ & 0.1 \\
       \bottomrule[0.5pt]\bottomrule[0.5pt]
       
    \end{tabular}
    }
    \vspace{-12pt}
    \label{params}
\end{table}
\begin{table}[ht]
    \centering
    \caption{Training Time for Different Training Method of Each Epoch}
    \resizebox{0.99\columnwidth}{!}{
    \begin{tabular}{c|c|c|c|c|c|c}
    \toprule[0.5pt]\toprule[0.5pt]
        Algorithm &  LOGNN & MLP & LOGNN(Sup) & MLP(Sup) & LOGNN(AC) & MLP(AC) \\ \hline
        Train Time(s) & 0.9308 & 0.7775 & 218.7574 & 218.6094 & 1.1656 & 0.8952 \\ 
    \toprule[0.5pt]\toprule[0.5pt]
    
    \end{tabular}}
    \vspace{-15pt}
    \label{tab-train}
\end{table}

\begin{table*}[ht]
    \centering
    \setlength{\tabcolsep}{1.15mm}{} 
    \renewcommand{\arraystretch}{0.76}
    \newcolumntype{P}[1]{>{\centering\arraybackslash}p{#1}}
    \caption{Total Tasks Processing Delay of Different Algorithms}
    \resizebox{1.99\columnwidth}{!}{
    \begin{tabular}{c|c|c|c|c|c|c|c|c|c|c|c|c|c|c|c|c|c|c}
    \toprule[0.5pt]\toprule[0.5pt]
        \begin{normalsize}
    		\diagbox[width=1.5cm,height=1cm,innerleftsep=0mm,innerrightsep=3mm]{Method}{$M$}
        \end{normalsize}
      & 2 & 3 & 4 & 5 & 6 & 7 & 15 & 16 & 17 & 18 & 19 & 20 & 25 & 26 & 27 & 28 & 29 & 30\\ \hline 
      LOGNN &6.48 & 7.80 & \textbf{6.08}  &  \textbf{6.39}  &  \textbf{7.13}  &  \textbf{6.08} & \textbf{5.15} & \textbf{5.32} & \textbf{5.50} & \textbf{5.59} & \textbf{5.27} & \textbf{5.15} & \textbf{5.01} &	\textbf{5.23} &	\textbf{5.29} &	\textbf{5.01} &	\textbf{5.11} &	\textbf{5.17}\\ \hline 
      MLP(DI) & 21.52 & 71.19 & 164.15 & 230.58 & 432.30 & 348.60 & 230.64  & 236.89 & 101.10 & 123.28 & 263.58 & 389.97 & 340.43 &	5162.02 &	4108.91 &	4912.52 &	5396.26 &	5444.65 \\ \hline 
      MLP(TR) & 10.2376 & 12.34 & 8.85 & 10.76 & 10.98 & 8.52 & 10.10 & 7.59 & 7.17 & 7.21 & 7.96 & 6.17 & 6.95  &  6.95  &  6.94  &  5.57  &  7.47  &  6.19 \\ \hline 
      GA & \textbf{5.68}  &  \textbf{7.07} &  11.51 &  16.53 & 20.72 & 25.45 &96.13 & 110.38 & 116.72 & 119.81 & 135.23 & 138.57& 196.73 &	228.13 &	248.36 & 226.52 & 277.60 &	279.42 \\ 
       \bottomrule[0.5pt]\bottomrule[0.5pt]
       
    \end{tabular}
    }
    \vspace{-12pt}
    \label{tab-avg}
\end{table*}
\begin{table*}[ht]
    \centering
    \setlength{\tabcolsep}{1.15mm}{} 
    \renewcommand{\arraystretch}{0.76}
    \newcolumntype{P}[1]{>{\centering\arraybackslash}p{#1}}
    \caption{Total Task Processing Delay (Algorithm Inference Delay Included) of Different Algorithms}
    \resizebox{1.99\columnwidth}{!}{
    \begin{tabular}{c|c|c|c|c|c|c|c|c|c|c|c|c|c|c|c|c|c|c}
    \toprule[0.5pt]\toprule[0.5pt]
        \begin{normalsize}
    		\diagbox[width=1.5cm,height=1.0cm,innerleftsep=0mm,innerrightsep=3mm]{Method}{$M$}
        \end{normalsize}
      & 2 & 3 & 4 & 5 & 6 & 7 & 15 & 16 & 17 & 18 & 19 & 20 & 25 & 26 & 27 & 28 & 29 & 30\\ \hline 
      LOGNN & \textbf{6.48} &\textbf{7.80} &\textbf{6.09} &\textbf{6.39} &\textbf{7.14} &\textbf{6.08} &\textbf{5.16} &\textbf{5.34} &\textbf{5.51} &\textbf{5.62} &\textbf{5.28} &\textbf{5.17} &\textbf{5.03} &\textbf{5.25} &\textbf{5.30} &\textbf{5.03} &\textbf{5.12} &\textbf{5.18}\\ \hline 
      MLP(DI) & 21.52 &71.19 &164.15 &230.58 &432.30 &348.60 &230.64 &236.90 &101.10 &123.29 &263.58 &389.97 &340.43 &5163 &4109 &4913 &5397  &5445  \\ \hline 
      MLP(TR) & 13.30 &15.38 &11.89 &13.57 &13.78 &11.29 &12.42 &9.88 &9.47 &9.59 &15.40 &8.24 &9.83 &10.31 &9.83 &13.73 &10.57 &9.48  \\ \hline 
      GA & 31.97 &50.91 &38.54 &31.38 &31.47 &36.07 &106.01 &119.01 &125.34 &128.41 &145.08 &145.63 &219.68 &255.79 &272.93 &237.48 &283.04 &284.96  \\ 
       \bottomrule[0.5pt]\bottomrule[0.5pt]
       
    \end{tabular}
    }
    \vspace{-14pt}
    \label{tab-time}
\end{table*}

\subsection{Comparison on Convergence of Different Training Methods}

To effectively evaluate the proposed unsupervised training method, we employ three different methods to train the LOGNN and MLP models. These methods include the default unsupervised learning method, the supervised learning method, and the actor-critic-based RL training method \cite{8650160}. The labels in supervised learning for LOGNN(Sup) and MLP(Sup) are obtained by the GA method. In contrast, the RL training method employs an actor net to optimize resource allocation, while a critic net provides the update gradient for the actor net. As label generation latency is considered as part of the training delay, the supervised training method has the largest training delay for each training epoch, as illustrated in Table \ref{tab-train}. Additionally, the supervised training method has worse convergence performance than the proposed unsupervised training method on the training set, as shown in Fig.~\ref{fig-con}. This can be attributed to the fact that the performance of an NN trained using supervised learning depends heavily on the quality of the training labels. Since the GA cannot precisely provide optimal solutions for supervised labels, the performance of supervised training cannot be guaranteed. Moreover, the supervised training method has a larger performance gap than the unsupervised counterpart on the test set due to the fact that the supervised training method is easy to overfit the training set.

Despite its faster training speed in one epoch, the RL-based method fails to converge. This is due to that the RL-based method requires the joint training of two NNs, with the performance of the critic network determining the accuracy of the updated direction of the actor net. Moreover, the output of the actor net determines the input distribution of the critic net. Therefore, even a slight perturbation can lead to non-convergence of the RL training, as evidenced by the results in Table \ref{tab-train} and Fig.~\ref{fig-con}. The proposed unsupervised training method outperforms supervised and RL training methods both in training speed and convergence speed, which suggests that when the environment changes dramatically, the unsupervised method can be leveraged to fine-tune the NN with low training costs, even in devices with limited computing resources.

\subsection{Performance Evaluation on Scalability}
Table \ref{tab-avg} shows the performance of diverse methods across varying numbers of edge servers and users, with user count $N$ consistently double that of server count $M$. Owing to MLP's lack of scalability, directly employing a fixed-size MLP with unaltered parameters only optimizes resource allocation for a subset of edge servers and users, with the remainder, allocated randomly. Consequently, MLP(DI)'s performance proves unstable and consistently inferior to alternative methods. In contrast, LOGNN consistently delivers optimal results, barring instances where $M=2,3$, in which GA outperforms LOGNN. As a near-brute search optimization method, GA can locate the optimal solution through an exhaustive evolutionary search when optimization problem complexity remains low with small $M$ values. However, as the problem size expands, GA struggles to pinpoint the objective function's optimal value, whereas LOGNN more effectively extracts MEC features to inform better decision-making.

It is important to note that LOGNN's performance dips when $M$ values are low but stabilizes when $M > 15$, exhibiting minimal fluctuations. GNN is better suited for extracting graph structure features rather than individual node features. For smaller $M$ values, edge node features like user task size $d_{i}$ and server computational resources $f_{j}^{s}$ substantially influence the optimal solution. Yet, as $M$ increases, cooperation and interference between graph nodes bear greater weight on optimization problem solutions. Given GNN's proficiency in extracting graph structural features, it achieves superior performance. Furthermore, LOGNN's output solutions maintain stable and similar objective values, as graph structures exert a more significant influence on problems than specific node features.

Notably, LOGNN consistently outperforms MLP-based approaches, even when MLP(TR) employs an architecture tailored for specific server and user counts and is trained for that particular MEC scale. This highlights that, for a controller overseeing a dynamic MEC, utilizing the proposed LOGNN directly addresses changes in user and server numbers, ensuring high performance without resorting to time-consuming training or fine-tuning.

\subsection{Performance Evaluation on Computational Efficiency}
Previous research works on MEC usually divide the processing delay of edge tasks into two main components, i.e., transmission delay and computation delay. However, task processing delay more generally refers to the time between task generation and completion. Thus, the inference delay of the resource allocation algorithm should be considered as a third component of task processing delay. To provide a comprehensive evaluation of different algorithms, we compare the task processing delays comprising algorithm inference delay, transmission delay, and computing delay in Table \ref{tab-time}. Although the sum of task transmission and computing delay optimized by GA is smaller than LOGNN when $M$ is less than 4, the inference delay of GA is remarkably high. As shown in Table \ref{tab-avg}, the summing delay of task processing and algorithm inference of GA is consistently much larger than LOGNN. Therefore, due to the high performance and small inference delay, the proposed LOGNN achieves the best performance among all methods. This highlights the potential of LOGNN in optimizing real-time resource allocation in MEC, offering valuable insights into its efficacy.

\section{Conclusion}
In this paper, we have proposed a LOGNN-based resource management method for MEC systems to address the scalable challenges in both the dimension of input space and decision space that are caused by varying numbers of servers and users. Furthermore, we have exploited a label-free unsupervised training method to reduce the training cost of LOGNN. Simulation results have demonstrated that the unsupervised LOGNN can efficiently and flexibly optimize the task offloading and resource allocation in MEC with an arbitrary number of servers and users with high performance and convergence speed. By implementing this scheme in MEC, we can effectively handle dynamic changes in the number of users and servers while significantly reducing edge task processing delay. For future research, we will study how to use the pre-trained GNN to generally optimize the MEC without fine-tuning.

\section*{Acknowledge}
This work was supported by the National Key Research and Development Program of China (2020YFB1807700), and the National Natural Science Foundation of China (NSFC) under Grant No. 62071356 and No. 62201414, and the fundamental research funds for the central universities under grant ZYTS23175.

\bibliography{ref}
\bibliographystyle{IEEEtran}

\ifCLASSOPTIONcaptionsoff
  \newpage
\fi

\end{document}